\documentclass[amsmath,amssymb,aps,floats,amsfonts,notitlepage,superscriptaddress,eqsecnum,nofootinbib,prd,a4paper,twocolumn,longbibliography]{revtex4-1}

\usepackage[utf8]{inputenc}
\usepackage[]{graphicx}
\usepackage{hyperref}
\usepackage{url}
\usepackage{color}
\usepackage[usenames,dvipsnames,svgnames,table]{xcolor}
\usepackage{multirow}
\allowdisplaybreaks[1]

\newcommand{\Rin}{{R^{\text{in}}_{\ell m \omega}}}
\newcommand{\Rup}{{R^{\text{up}}_{\ell m \omega}}}

\newcommand{\XtH}{{X_2{}^{\!1/2}}}

\newcommand{\Slm}{ S_{\ell m}}
\newcommand{\Ylm}{Y_{\ell m}}
\newcommand{\Uhyp}{\mathop{U\!}\nolimits}

\newcommand{\eig}{\lambda_{\ell m}}

\newcommand{\XinMST}{{X^{\text{in(MST)}}_{\ell m}}}
\newcommand{\XupMST}{{X^{\text{up(MST)}}_{\ell m}}}

\newcommand{\diff}[2]  {\frac{d #1}{d #2}}

\newcommand{\pdiff}[2]  {\frac{\partial #1}{\partial #2}}
\newcommand{\spdiff}[2] {\frac{\partial^2 #1}{\partial #2^2}}

\begin{document}

\title{Analytical high-order post-Newtonian expansions for spinning extreme mass ratio binaries}

\author{Chris Kavanagh}
\affiliation{School of Mathematics and Statistics and Complex \& Adaptive Systems Laboratory, University College Dublin, Belfield, Dublin 4, Ireland.}

\author{Adrian C.~Ottewill}
\affiliation{School of Mathematics and Statistics and Complex \& Adaptive Systems Laboratory, University College Dublin, Belfield, Dublin 4, Ireland.}

\author{Barry Wardell}
\affiliation{School of Mathematics and Statistics and Complex \& Adaptive Systems Laboratory, University College Dublin, Belfield, Dublin 4, Ireland.}
\affiliation{Department of Astronomy, Cornell University, Ithaca, NY 14853, USA}

\begin{abstract}
We present an analytic computation of Detweiler's redshift invariant for a point mass in a circular
orbit around a Kerr black hole, giving results up to 8.5 post-Newtonian order
while making no assumptions on the magnitude of the spin of the black hole. Our calculation is
based on the functional series method of Mano, Suzuki and Takasugi, and employs a rigorous mode-sum
regularization prescription based on the Detweiler-Whiting singular-regular decomposition. The
approximations used in our approach are minimal; we use the standard self-force expansion to linear
order in the mass ratio, and the standard post-Newtonian expansion in the separation of the binary.
A key advantage of this approach is that it produces expressions that include contributions at all
orders in the spin of the Kerr black hole. While this work applies the method to the specific case
of Detweiler's redshift invariant, it can be readily extended to other gauge invariant quantities
and to higher post-Newtonian orders.
\end{abstract}

\maketitle

\section{Introduction}
\label{sec:intro}

Binary black hole systems have been identified as one of the primary sources of gravitational waves
for current and planned generations of gravitational-wave detectors
\cite{LIGO,eLISA,eLISA:GWNotes}. Accurate models for the waveforms produced by gravitational-wave
sources are a crucial component in the data-analysis pipeline used to extraction of information
from gravitational wave observations. In the context of compact-object binary systems, the
production of waveform models typically relies on one of three fundamental methods: numerical
relativity (NR) simulations, post-Newtonian (PN) approximations, or gravitational self-force (GSF)
calculations using black hole perturbation theory.

Gravitational self-force calculations --- involving, for example, a solar mass black hole or
neutron star of mass $m$ in orbit around a massive black hole of mass $M$ --- are based on a
perturbative expansion of Einstein's equations, with the mass ratio $m/M$ as a small expansion
parameter. These so-called extreme mass ratio inspiral (EMRI) systems are well approximated by an
expansion to linear order in $m/M$. An alternative approach to the two-body problem --- valid when
the constituents are far apart --- is the post-Newtonian approximation, which expands the Einstein
equations in $v^2/c^2$, where $v$ is a representative velocity and $c$ is the speed of light. In
the context of binary systems, the post-Newtonian expansion maps onto an expansion in $1/r$, where
$r$ is the separation of the two objects.

The problem of gauge freedom in general relativity is a constant source of
difficulty in extracting meaningful information from calculations. It is often difficult to know
how much of the difference between two different results is due to merely a difference in choice of
gauge. The computation of gauge-invariant quantities provides a robust solution to this problem.
Independent of the choice of gauge, the computation of gauge-invariant quantities should agree
among different methods; any discrepancy between results can be confidently associated with errors
in the method. Even better, a gauge-invariant quantity computed within one approach can often be
used to inform --- and even derive results within --- other approaches to the same problem.

The strategy of making comparisons based on gauge invariant quantities has proven to be a fruitful
method for cross-pollination of results between NR, PN, and GSF
\cite{Blanchet:2009sd,Barack:2009ey,Damour:2009sm,Blanchet:2010cx,Blanchet:2010zd,Barack:2010ny,Dolan:2013roa,Dolan:2014pja,LeTiec:2011bk,Bernuzzi:2014owa,Damour:2016abl}.
This work continues a programme
\cite{Dolan:2013roa,Dolan:2014pja,Kavanagh:2015lva,Nolan:2015vpa,Hopper:2015icj} to compute the
gauge-invariant quantities that enable these valuable cross-comparisons. Our key new development is
the incorporation of spin effects, by applying methods similar to those of \cite{Kavanagh:2015lva}
to the case of analytic GSF calculations in Kerr spacetime, allowing us to add important
spin-dependent terms into PN and effective-one-body (EOB) \cite{Buonanno:1998gg} models. Given that most (if not all) astrophysical black holes
are expected to be spinning \cite{McClintock:2013vwa,Reynolds:2013qqa}, these additional terms are
crucial for faithfully representing the type of systems we expect gravitational wave detectors to
observe.

A parallel effort by Bini, Damour and Geralico \cite{Bini:2015xua} has recently been successful in
computing spin-dependent contributions to GSF-PN results --- that is, approximations to the
spacetime of a binary system using simultaneous expansions in the mass ratio and the binary
separation --- using a small-spin approximation. Our work provides two key advantages over these
results:
\begin{enumerate}
\item The only expansion our method relies only is the standard PN expansion in the
      inverse separation, $y\sim 1/r$ (in addition to the standard expansion in mass-ratio used in
      all self-force calculations). Importantly, this allows to provide for the first time
      expressions that are exact in the spin, $a$, of the larger black hole. The validity of our
      expressions in the high-spin regime is particularly important given observational evidence
      for black holes with near-extremal ($a > 0.95M$) spins
      \cite{McClintock:2013vwa,Reynolds:2013qqa}.
\item Our regularization procedure is based on an independently obtained expression
      for spheroidal-harmonic mode-sum regularization parameters derived from the
      Detweiler-Whiting singular field.
\end{enumerate}
In addition, our results provide a valuable independent check; indeed, they identify what we believe
to be an error in some terms in the order $y^{9.5}$ PN coefficient for $\Delta U$ given by
Ref.~\cite{Bini:2015xua}. This conclusion has been confirmed by an independent check against
numerically derived values \cite{Abhay}.

This paper follows the conventions of Misner, Thorne and Wheeler
\cite{Misner:1974qy}; a ``mostly positive'' metric signature, $(-,+,+,+)$, is
used for the spacetime metric, the connection coefficients are defined by
$\Gamma^{\lambda}_{\mu\nu}=\frac{1}{2}g^{\lambda\sigma}(g_{\sigma\mu,\nu}
+g_{\sigma\nu,\mu}-g_{\mu\nu,\sigma}$), the Riemann tensor is
$R^{\alpha}{}_{\!\lambda\mu\nu}=\Gamma^{\alpha}_{\lambda\nu,\mu}
-\Gamma^{\alpha}_{\lambda\mu,\nu}
+\Gamma^{\alpha}_{\sigma\mu}\Gamma^{\sigma}_{\lambda\nu}
-\Gamma^{\alpha}_{\sigma\nu}\Gamma^{\sigma}_{\lambda\mu}$, the Ricci tensor and
scalar are $R_{\alpha\beta}=R^{\mu}{}_{\!\alpha\mu\beta}$ and
$R=R_{\alpha}{}^{\!\alpha}$, and the Einstein equations are
$G_{\alpha\beta}=R_{\alpha\beta}-\frac{1}{2}g_{\alpha\beta}R=8\pi
T_{\alpha\beta}$. Standard geometrized units are used, with $c=G=1$, but we
include the explicit dependence on $G$ and $c$ in our post-Newtonian expansions
in cases where they are convenient for post-Newtonian order counting. We use
the spherical Boyer-Lindquist coordinates $\{t,r,\theta,\phi\}$ for the background Kerr
spacetime and write tensors in terms of these coordinate components.

\section{Redshift for circular geodesics in Kerr spacetime}

\subsection{Circular, equatorial geodesic orbits in Kerr spacetime}
In this work we are interested in the case of a point mass on a circular equatorial geodesic
in Kerr spacetime. In Boyer-Lindquist coordinates, the line element for a Kerr
black hole of mass $M$ and spin parameter $a$ is given by
\begin{align}
ds^2 = &
- \Big(1 - \frac{2 M r}{\Sigma}\Big) dt^2 + \frac{\Sigma}{\Delta} dr^2 + \Sigma d \theta^2 \nonumber \\ &
+ \Big( r^2 + a^2 + \frac{2 M r a^2 \sin^2 \theta}{\Sigma} \Big)\sin^2 \theta d \varphi^2\nonumber \\ & - \frac{4 a M r \sin^2 \theta}{\Sigma} dt d \varphi,
\end{align}
with
\begin{align}
\Sigma \equiv r^2 + a^2 \cos^2 \theta,
\quad \quad
\Delta \equiv r^2 - 2M r + a^2.
\end{align}
Circular equatorial orbits can be parametrized by the orbital frequency,
$\Omega$, which is related to the Boyer-Lindquist radius of the orbit, $r_0$, by
\begin{equation}
\label{eq:Omega}
  \Omega = \dfrac{d\varphi}{dt} = \frac{ M^{1/2}}{r_0^{3/2}+a M^{1/2}}.
\end{equation}
Adopting the convention that $u^\varphi$ (and hence the orbital angular momentum) is
always positive, the orbital angular momentum and energy per unit mass for such orbits
are given by
\begin{align}
\mathcal{L} &= \frac{M (a^2 + r_0^2 - 2 a \sqrt{M r_0})}{\sqrt{
 M r_0} \sqrt{r_0^2 - 3 M r_0 + 2 a \sqrt{M r_0}}} = u_\varphi, \\
 E &= \frac{r_0^2 - 2M r_0 + a \sqrt{M r_0}}{r_0\sqrt{r_0^2 - 3M r_0 + 2a \sqrt{M r_0}}} = - u_t.
\end{align}
Within this convention, prograde and retrograde orbits are distinguished by the sign of
$a$: $a > 0$ for the former; $a < 0$ for the latter.

\subsection{Redshift invariant}
The key result of this work is the computation of a post-Newtonian expansion of
Detweiler's redshift invariant \cite{Detweiler:2008ft} for circular equatorial orbits in Kerr spacetime.
In this context, the redshift invariant is the constant of proportionality between
the particle's 4-velocity, $u^\alpha$, and the helical Killing vector of the system,
$k^\alpha$, i.e.
\begin{equation}
  u^\alpha = U k^\alpha.
\end{equation}
In the background spacetime, $U$ is equal to the time component of the 4-velocity,
$u^t$. Provided $k^\alpha$ is a Killing vector of both the background and perturbed
spacetime (i.e. they both share the same helical symmetry) the contribution to $U$
from the linear order metric contribution is simply given by
\begin{equation}
  \Delta U = \frac12 h_{\alpha \beta} u^\alpha u^\beta u^t.
\end{equation}
This is invariant in the sense that it does not change under helically symmetric gauge
transformations that also respect the helical symmetry of the worldline. When making comparisons between calculations using different gauges, care must be taken to ensure that these criterion are satisfied, or that the appropriate transformation of $\Delta U$ is accounted for \cite{Sago:2008id}.

\section{Regularised metric perturbations on a Kerr background spacetime}

\subsection{Perturbations of Kerr spacetime}
In the Kerr spacetime Teukolsky showed, using the Newman-Penrose formalism, that the dynamics of a metric perturbation are described by the
evolution of tetrad components of the Weyl tensor \cite{Teukolsky:1972my,Teukolsky:1973ha}. This requires choosing a set of four null vectors $e^\alpha_i$, $i=1,...4$, two real and two a complex conjugate pair.
Using Boyer-Lindquist coordinates, a particular set of null vectors --- the Kinnersly tetrad --- is
\begin{align*}
e^\alpha_1&=l^\alpha=\frac{1}{\Delta}(r^2+a^2,\Delta,0,a), \\
e^\alpha_2&=n^\alpha=\frac{1}{2 \Sigma}(r^2+a^2,-\Delta,0,a) \\
e^\alpha_3&=m^\alpha=-\frac{\bar{\varrho}}{\sqrt{2}}(i a \sin \theta,0,1,\frac{i}{\sin \theta}) \\
e^\alpha_4&=\bar{m}^\alpha=-\frac{\varrho}{\sqrt{2}}(-i a \sin \theta,0,1,\frac{-i}{\sin \theta}) 
\end{align*}
Using this tetrad, five of the spin-coefficients vanish, leaving us with 
\begin{align*}
\varrho=\frac{-1}{r-i a \cos \theta},
\qquad\tau=\frac{-i a \sin \theta}{\sqrt{2} \Sigma} ,
\qquad\beta=-\frac{\bar{\varrho}\cot \theta}{2\sqrt{2}},\\
\gamma=\mu+\frac{r-M}{2 \Sigma},
\qquad \mu=\frac{\Delta \varrho}{2 \Sigma},
\qquad \varpi=\frac{i a\varrho^2 \sin \theta }{\sqrt{2}},
\end{align*}
\begin{align*}
\alpha=\varpi-\bar{\beta} .
\end{align*}
The perturbed Weyl tensor $C_{\alpha\beta\gamma\delta}$, then has two components of interest 
\begin{align}
\psi_0&=-C_{\alpha\beta\gamma\delta}l^\alpha m^\beta l^\gamma m^\delta, \\
\psi_4&=-C_{\alpha\beta\gamma\delta}n^\alpha \bar{m}^\beta n^\gamma \bar{m}^\delta,
\end{align}
which both independently contain all of the radiative information about perturbations of Kerr
spacetime. Teukolsky's key insight was that the equations for $\psi_0$ and $\psi_4$ decouple, and
each satisfy a separable PDE now known as the Teukolsky equation:
\begin{widetext}
\begin{align}
\left[\frac{(r^2+a^2)^2}{\Delta}-a^2\sin^2\theta\right]\spdiff{\psi}{t}+\frac{4 M a r}{\Delta}&\frac{\partial^2 \psi}{\partial t \partial \varphi}+
\left[\frac{a^2}{\Delta}-\frac{1}{\sin^2\theta}\right]\spdiff{\psi}{\varphi} 
-\Delta^{-s}\pdiff{}{r}\left(\Delta^{s+1}\pdiff{\psi}{r}\right) -\frac{1}{\sin \theta}\pdiff{}{\theta}\left(\sin\theta \pdiff{\psi}{\theta}\right)  \nonumber\\
-2s\left[\frac{a(r-M)}{\Delta}+\frac{i \cos\theta}{\sin^2\theta}\right]&\pdiff{\psi}{\varphi}
-2 s \left[\frac{M(r^2-a^2)}{\Delta}-r -i a \cos\theta\right]\pdiff{\psi}{t}+(s^2\cot^2\theta-s)\psi = 4 \pi \Sigma T,
\label{Eq:TeukMaster}
\end{align} 
\end{widetext}
where 
\begin{equation*}
\psi=\psi_0 \implies T = 2\, T_0
\end{equation*}
and
\begin{equation*}
\psi=\varrho^{-4} \psi_4 \implies T = 2 \varrho^{-4}\, T_4,
\end{equation*}
respectively.
In terms of tetrad components of the stress tensor, $T_{\textbf{ij}}=T^{\mu\nu}e_\mu^ie_\nu^j$, the
source terms for the Teukolsky equation are
\begin{align}
T_0&=(\boldsymbol{\delta}+\bar{\varpi}-\bar{\alpha}-3 \beta-4 \tau) \times\nonumber\\
& \qquad \left[(\boldsymbol{D}-2\epsilon-2\bar{\varrho})T_{13} -(\boldsymbol{\delta}+\bar{\varpi}-2\bar{\alpha}-2 \beta)T_{11}\right] \nonumber \\
& +(\boldsymbol{D}-3\epsilon+\bar{\epsilon}-4\varrho-\bar{\varrho}) \times \nonumber \\
& \qquad\left[(\boldsymbol{\delta}+2 \bar{\varpi} -2\beta)T_{13}-(\boldsymbol{D}-2\epsilon+2\bar{\epsilon}-2\bar{\varrho})T_{33}\right], \\
\nonumber\\
T_4&=(\boldsymbol{\Delta}+3\gamma-\bar{\gamma}+4 \mu+\bar{\mu}) \times \nonumber \\
& \qquad \left[(\boldsymbol{\bar{\delta}}-2\bar{\tau}+2\alpha)T_{24}
  -(\boldsymbol{\Delta} +2\gamma-2\bar{\gamma}+\bar{\mu})T_{44}\right] \nonumber \\
& + (\boldsymbol{\bar{\delta}}-\bar{\tau}+\bar{\beta}+3\alpha+4\varpi) \times \nonumber \\
& \qquad \left[(\boldsymbol{\Delta}+2 \gamma +2\bar{\mu})T_{24}-(\boldsymbol{\bar{\delta}}-\bar{\tau}+2\bar{\beta}+2\alpha)T_{22}\right],
\end{align}
where $\boldsymbol{D}=l^\mu\partial_\mu$, $\boldsymbol{\Delta}=n^{\mu}\partial_\mu$ and $\boldsymbol{\delta}=m^\mu\partial_\mu $. In this work we are interested in a perturbation sourced by a point particle,
\begin{align}
T^{\mu\nu}=m_p \frac{u^\mu u^\nu}{u_t \Sigma_0 \sin \theta_0}\delta(r-r_0)\delta(\theta-\theta_0)\delta(\varphi-\varphi_0)
\end{align}
where a subscript-$0$ denotes evaluation on the worldline. In particular, for circular orbits $\theta_0=\tfrac{\pi}{2}$, $\varphi_0=\Omega t$, so that $T_{\textbf{ij}}$ are the tetrad components of the stress tensor with the tetrad vectors evaluated at the position of the particle. 
The Teukolsky master equation, Eq.~\eqref{Eq:TeukMaster}, may be separated by writing
\begin{align}
\psi=\sum_{\ell,m}\int e^{-i\omega t}{}_s\Slm(\theta,\varphi; a \omega)\, {}_s R_{\ell m \omega}(r) d\omega.
\end{align}
Here, ${}_s\Slm (\theta, \varphi; a \omega)$ are the oblate spin-weighted spheroidal harmonics
with defining equation
\begin{align}
&\bigg[\frac{1}{\sin \theta}\diff{}{\theta}\left(\sin\theta \diff{  }{\theta}\right)-a^2 \omega^2 \sin^2\theta
-\frac{(m+s\cos\theta)^2}{\sin^2\theta} \nonumber \\
&  \quad -2 a \omega s \cos\theta+s+2 m a \omega + {}_s \eig\bigg]{}_s\Slm (\theta, \varphi; a \omega)=0. \label{Eq:Slmeq} 
\end{align}
The spin-weighted spheroidal harmonics are orthonormal on the two-sphere,
\begin{equation}
\label{eq:spheroidal-normalisation}
\int {}_s\Slm(\theta, \varphi; a \omega) {}_s S^{\ast}_{\ell' m'}(\theta, \varphi; a \omega)  {\rm d} \Omega = \delta_{\ell \ell'} \delta_{m m'},
\end{equation}
so that in the limit $a\omega\rightarrow 0$ they coincide with the standard spin-weighted spherical
harmonics ${}_s\Ylm(\theta, \varphi)$ (this normalisation is consistent with the Meixner-Sch\"afke
convention for the spheroidal Legendre functions, $S^{(1)}_{m \ell}(a \omega, \cos \theta)$,
defined in \cite{Abramowitz:Stegun}). The radial functions are solutions to
\begin{align}
\bigg[\Delta^{-s}\diff{}{r}\left(\Delta^{s+1}\diff{}{r}\right)+&\frac{K^2-2 i s (r-M)K}{\Delta}\nonumber \\
+4 i s \omega r -{}_s &\eig\bigg]{}_s R_{\ell m \omega}(r)
={}_sT_{\ell m \omega}, \label{Eq:RadTeuk} 
\end{align}
where $K = (r^2+a^2)\omega - a m$, and where the source term on the right hand side is obtained from a mode decomposition of $T_0$ or
$T_4$. In the next section, we will develop analytic expressions for the appropriate homogeneous
solutions to this equation.

\subsection{Homogeneous Solutions of the Teukolsky equation}
\label{sec:methods}

We now seek solutions to Eq.~\eqref{Eq:RadTeuk} for $s=\pm2$, $\ell \ge |s|$, $-\ell \le m \le
\ell$ and $\omega \in \mathbb{R}$. To this end, we build up a complete set of $\ell$ modes from
three distinct sections. For all $\ell\geq 2$ we can use the solutions to the Teukolsky equation
give by Mano, Suzuki and Takasugi, as detailed in a review article by Sasaki and Tagoshi
\cite{Sasaki:2003xr}. Furthermore, for sufficiently large values of $\ell$ the regularity of these
solutions can be exploited with an ansatz to produce general expressions for arbitrary (large) $\ell$
and $m$. As such, in practice we: (i) compute a finite number of specific-$\ell$ values (with the
exact number of required specific-$\ell$ values needed depending on the final PN order desired)
using the MST solutions; and (ii) obtain the remaining modes from our ansatz. The third section of
the solution relates to the non-radiative modes, and is addressed in Sec.~\ref{sec:completion}.

\subsubsection{Low \texorpdfstring{$\ell$}{l} modes: MST}

Descriptions of the construction of the MST series solutions to the Teukolsky equation are widely available in the literature \cite{Sasaki:2003xr}, so we will give here only a brief overview. 

In computing PN expansions, we find two natural small parameters; the frequency $\omega=m \Omega$
and the inverse of the radius, $1/r$, which is related to the orbital frequency, $\Omega$, via
Eq.~\eqref{eq:Omega}. This double expansion is handled indirectly by expanding in the inverse of
the speed of light, $1/c$, and introducing the auxiliary variables $X_1=G M/r$ and
$X_2^{1/2}=\omega r$. Dimensional analysis shows that each of these variables carry a factor of
$1/c$, and are of the same order in the large-$r$ limit. Towards the end of our calculations we
will change back to a single expansion variable, namely $y= (M \Omega)^{2/3}$ and use this relation
to define $X_1(y)$ and $X_2(y)$.

\subsubsection*{The MST expansion of \texorpdfstring{${}_s \Rin(r)$}{Rin(r)}}

The homogeneous solution satisfying retarded boundary conditions at the horizon can be written as a convergent infinite sum of hypergeometric functions
\begin{align}
\label{STeq120}
&{}_s \Rin (r) = C_{({\rm in})}^{\nu} (x) \sum_{n=-\infty}^{\infty} a_n^{\nu} \times \nonumber\\
&\quad{}_2F_1(n+1+\nu -i\tau, -n-\nu-i\tau, 1-s-i \epsilon-i \tau , x ) \, , 
\end{align}
with
\begin{align*}
C_{(\text{in})}^{\nu} (x) &=e^{i\epsilon \kappa x}(-x)^{-s-i(\epsilon+\tau)/2}
(1-x)^{i(\epsilon-\tau)/2},
\end{align*}
where
\begin{align*}
x&= \frac{(r_+- r)c^2}{2G M\kappa} = 
\frac{2\kappa\eta^2 X_1}{1+(\kappa-1) \eta^2 X_1},\\
\epsilon&= 2GM\omega/c^3 = 2 X_1 X_2{}^{1/2} \eta^3,
\end{align*}
and where
\begin{align*}
 q&=a/M \qquad \qquad \qquad \, \kappa=\sqrt{1-q^2}\nonumber \\
r_\pm&=GM(1\pm\kappa)/c^2\qquad \tau=(\epsilon - m q)/\kappa.
\end{align*}
Note that $\tau$ is $O(\epsilon)$ in Schwarzschild and also for $m=0$ in Kerr, but is $O(1)$ for $m\neq 0$ in Kerr.
Here $\nu$ is the well-known ``renormalised angular momentum''; for the purpose of this section the critical feature is that
$\nu=l+O(\epsilon^2)$.
We may write $C_{(\text{in})}^{\nu} (x)$ as
\begin{equation*}
C_{(\text{in})}^{\nu} (x) =e^{i\epsilon \kappa x}(2\kappa\eta^2 X_1)^{s+i\tau}
\frac{\left[ 1+(\kappa-1) \eta^2 X_1\right]^{i\frac{\epsilon-\tau}{2}}}{\left[1 - (1+\kappa)\eta^2X_1\right]^{s+i\frac{\epsilon+\tau}{2}}},
\end{equation*}
and more convenient expressions for they hypergeometric function can be obtained by using the
decomposition
\begin{align}
\label{eq:F1F2}
{}_2 & F_1(a,b,c;\zeta) =
\nonumber \\
&\, \frac{\Gamma(c) \Gamma(b-a)}{\Gamma(b) \Gamma(c-a)} (1-\zeta)^{-a} 
{}_2 F_1(a,c-b,a-b+1,\tfrac{1}{1-\zeta})
\nonumber \\
&\,+ \frac{\Gamma(c) \Gamma(a-b)}{\Gamma(a) \Gamma(c-b)} (1-\zeta)^{-b} {}_2 F_1(c-a,b,b-a+1,\tfrac{1}{1-\zeta}).
\end{align}
We denote the first and second terms on the right hand side by $F_1$ and $F_2$, respectively;
this leads to the $R^\text{in}=R_0^\nu+R_0^{-\nu-1}$ 
representation of Eq.~(137) in~\cite{Sasaki:2003xr} if we let $n\rightarrow-n$ in the $F_1$ sum.

High order expansions of these expressions can prove computationally quite expensive if done
without care. An examination of the leading order behaviour in $\eta=1/c$ of each term in the sums
involved in both $F_1$ and $F_2$ helps to minimise the expansion of each $n$-value and determine a
look-up table for truncating the infinite sum to a given order in $\eta$. In doing this one must
take care of the irregular behaviour of the the series coefficients $ a_n^\nu $, the $\Gamma$
functions, and the ${}_2 F_1 $ with changing $n$. We summarise our results in Tables~\ref{tab:a},
\ref{tab:F} and \ref{tab:aF}. One minor complication we find is that the static $m=0$ case must be
handled separately. In the circular orbit case this problem can be circumvented; when $m=0$ then
also $\omega=m \Omega=0$. Closed form analytic expressions for the solutions in this case were derived by Teukolsky \cite{Teukolsky:1974}, and more recently given in \cite{O'Sullivan:2014cba} and \cite{vandeMeent:2015lxa}.

\begin{table*}[htb]
\begin{center}
\begin{tabular}{|c|c|c|c|c|c|}
\hline
\hline
& $n\leq -2\ell-1$ & $-2\ell \leq n\leq -\ell-1$ & $n=-\ell$ & $n=-\ell+1$ &  $n\geq -\ell+2$  \\
\hline
$a_n (m\neq 0)$ &$3(|n|-2)$&$3|n|$ &  $3\ell+3$ &  $3\ell+3$&  $3|n|$ \\
$a_n (m = 0)$ &$3(|n|-1)$&$3(|n|+1)$ &  $3\ell+6$ &  $3\ell+3$&  $3|n|$ \\
\hline
\hline
\end{tabular}
\caption{\label{tab:a} The leading behaviour of the MST coefficients for the spin $-2$ Teukolsky equation in terms of powers of $\eta$.}
\end{center}
\end{table*}

\begin{table*}[htb]
\begin{center}
 \begin{tabular}{|c|c|c|c|c|c|c|}
\hline\hline
& $n\leq -\ell-3$ & $n=-\ell-2$& $n= -\ell-1$ & $n=-\ell$ & $n=-\ell+1$ &  $n\geq -\ell+2$  \\
\hline
$F_1 (m\neq 0)$&$2n+2\ell+2$&$-2$&$-4^\dagger$&$-4$&$-2$&$2n+2\ell-1$\\
$F_1 (m = 0)$&$2n+2\ell+2$&$-2$&$-1^\dagger$&$-1$&$1$&$2n+2\ell+2$\\
$F_2 (m\neq 0)$&$-2n-2\ell-3$&$-2$&$-4$&$-4^\dagger$&$-2$&$-2n-2\ell$\\
$F_2 (m = 0)$&$-2n-2\ell$&$1$&$-1$&$-1^\dagger$&$-2$&$-2n-2\ell$\\
\hline\hline
\end{tabular}
\caption{\label{tab:F} The leading behaviour of the hypergeometric functions appearing as the first
and second terms in Eq.~(\ref{eq:F1F2}). The terms marked with a $^\dagger$ highlight the interplay
between parameters and argument; see \cite{Kavanagh:2015a} for a full discussion.}
\end{center}
\end{table*}

\begin{table*}[htb]
\begin{center}
\begin{tabular}{|c|c|c|c|c|c|c|c|c|}
\hline\hline
& $n\leq -2\ell-1 $&$-2\ell\leq n\leq -\ell-3$ & $n=-\ell-2$  &  $n= -\ell-1$ & $n=-\ell$ & $n=-\ell+1$ &  $n\geq -\ell+2$  \\
\hline
$\eta^{2\ell+4}a_nC_{(\text{in})}^{\nu} F_1(m\neq0)$&$|n|+4\ell-4$&$|n|+4\ell+2$&$5\ell+4$&$5\ell-1^\dagger$&$5\ell-1$&$5\ell+1$&$3|n|+2n+4\ell-1$\\
$\eta^{2\ell+4}a_nC_{(\text{in})}^{\nu} F_1(m=0)$&$|n|+4\ell-1$&$|n|+4\ell+5$&$5\ell+7$&$5\ell+5^\dagger$&$5\ell+2$&$5\ell+4$&$3|n|+2n+4\ell+2$\\
$\eta^{2\ell+4}a_nC_{(\text{in})}^{\nu} F_2(m\neq0)$&$5|n|-9$&$5|n|-3$&$5\ell+4$&$5\ell-1$&$5\ell-1^\dagger$&$5\ell+1$&$3|n|-2n$\\
$\eta^{2\ell+4}a_nC_{(\text{in})}^{\nu} F_2(m=0)$&$5|n|-3$&$5|n|+3$&$5\ell+10$&$5\ell+5$&$5\ell+5^\dagger$&$5\ell+1$&$3|n|-2n$\\
\hline\hline
\end{tabular}
\caption{\label{tab:aF} The combined behaviour following from Tables~\ref{tab:a} and \ref{tab:F}. We take out a factor of $\eta^{2l+4}$
for normalisation fixing the largest term as $O(1)$,  everything higher can be read as relative.}
\end{center}
\end{table*}

\subsubsection*{The MST expansion of \texorpdfstring{${}_s \Rup(r)$}{Rup(r)}}

The homogeneous solution satisfying radiative boundary conditions at infinity can be written as a
sum over irregular confluent hypergeometric functions with the same series coefficients
\begin{align}
\label{eq:STeq153}
&{}_s \Rup (r) =
 C_{({\rm up})}^{\nu} (z) \frac{\Gamma(\nu-s+1+i\epsilon)}{\Gamma(\nu+s+1-i\epsilon)}\times \nonumber\\
&\, \sum_{n=-\infty}^{\infty}
a_n^{\nu}(2iz)^n  \tilde{U}(n+1+s+\nu-i\epsilon,2n+2+2\nu;-2iz),
\end{align}
where
\begin{align*}
\tilde{U}(a,b,\zeta) = \frac{\Gamma (a)}{\Gamma (a^*-2s) }U(a,b,\zeta),
\end{align*}
with
\begin{align*}
C_{({\rm up})}^{\nu} (z) &= 2^{\nu}e^{-\pi \epsilon}e^{-i\pi(\nu+1+s)}
e^{iz} \frac{z^{\nu+i(\epsilon+\tau)/2}}{(z-\epsilon\kappa)^{s+i(\epsilon+\tau)/2}},
\end{align*}
and 
\begin{align*}
z=\omega(r-r_-)=\epsilon\kappa(1-x) =\eta  X_2{}^{1/2}\bigl[1-(1-\kappa)\eta^2 X_1\bigr] .
\end{align*}
Then,
\begin{align*}
C_{({\rm up})}^{\nu} (z) &= 2^{\nu}e^{-\pi \epsilon}e^{-i\pi(\nu+1+s)}e^{iz}(\eta  X_2{}^{1/2})^{\nu-s} \times \\
&\qquad \frac{\bigl[1-(1-\kappa)\eta^2 X_1\bigr]^{\nu+i\frac{\epsilon+\tau}{2}}}{\bigl[1-(1+\kappa)\eta^2 X_1\bigr]^{s+i\frac{\epsilon+\tau}{2}}}.
\end{align*}
Note that the prefactors here have been taken to agree with Sasaki and Tagoshi~\cite{Sasaki:2003xr}.

Now using the standard identity
\begin{align*}
\Uhyp\left(a,b,z\right)&=\frac{\Gamma\left(1-b\right)}{\Gamma\left(a-b+1\right)}M\left(a,b,z\right)+\\
&\qquad\frac{\Gamma\left(b-1\right)}{\Gamma\left(a\right)}z^{1-b}M\left(a-b+1,2-b,z\right),
\end{align*}
we may split \eqref{eq:STeq153} into two more manageable pieces $\tilde{\Uhyp}\left(a,b,\zeta\right)=\tilde{\Uhyp}_1\left(a,b,\zeta\right)+\tilde{\Uhyp}_2\left(a,b,\zeta\right)$, where
\begin{align}
\label{eq:Uhyp}
&\tilde{\Uhyp}_1\left(n+1+s+\nu-i\epsilon,2n+2+2\nu;-2iz\right)=\nonumber\\
&\qquad (-1)^{n-s}\frac{\sin(\nu+i\epsilon)\pi}{\pi} \times \nonumber\\
&\qquad\qquad\Gamma(n+1+s+\nu-i\epsilon)\Gamma\left(- 2n-1-2\nu\right)\times\nonumber\\
&\qquad\qquad \qquad M\left(n+1+s+\nu-i\epsilon,2n+2+2\nu;-2iz\right),
\end{align}
\begin{align}
\label{eq:Uhyp2}
&\tilde{\Uhyp}_2\left(n+1+s+\nu-i\epsilon,2n+2+2\nu;-2iz\right)=\nonumber\\
&\qquad \frac{\Gamma\left(2n+1+2\nu\right)}{\Gamma(n+1-s+\nu+i\epsilon)}z^{- 2n-2\nu-1}\times\nonumber\\
&\qquad\qquad M\left(-n+s-\nu-i\epsilon,- 2n-2\nu ,-2iz\right).
\end{align}

The leading order behaviour in $\eta$ of each term in the sums involved in both $\tilde{\Uhyp}_1$ and $\tilde{\Uhyp}_2$ are summarised  in  Tables~\ref{tab:U} and \ref{tab:aU}.
Once again we treat the $m=0$ case differently. The tables are presented, however the analytic expressions given in \cite{vandeMeent:2015lxa} can 
alternatively be used.

\begin{table*}[htb]
\begin{center}
 \begin{tabular}{|c|c|c|c|c|c|c|}
\hline\hline
& $n\leq -\ell-3$ & $n=-\ell-2$& $n= -\ell-1$ & $n=-\ell$ & $n=-\ell+1$ &  $n\geq -\ell+2$  \\
\hline
$\tilde{\Uhyp}_1$&$0$&$0$&$-5^\dagger$&$-6$&$-6$&$-3$\\
$\tilde{\Uhyp}_2$&$-2n-2\ell-4$&$-3$&$-5$&$-6^\dagger$&$-3$&$-2n-2\ell-1$\\
\hline\hline
\end{tabular}
\caption{\label{tab:U} The leading behaviour of the confluent hypergeometric functions appearing in Eqs.~\eqref{eq:Uhyp} and \eqref{eq:Uhyp2}. The terms marked with a $^\dagger$ highlight the interplay between parameters and argument; see \cite{Kavanagh:2015a} for a full discussion.}
 \end{center}
\end{table*}

\begin{table*}[htb]
\begin{center}
\begin{tabular}{|c|c|c|c|c|c|c|c|}
\hline\hline
& $n\leq -2\ell-1 $&$-2\ell\leq n\leq -\ell-3$ & $n=-\ell-2$&  $n= -\ell-1$ & $n=-\ell$ & $n=-\ell+1$ &  $n\geq -\ell+2$  \\
\hline

$\eta^{\ell-1}a_n C_{({\rm up})}^{\nu}(-2iz)^n\tilde{\Uhyp}_1 (m\neq 0)$&$2|n|+2\ell-5$&$2|n|+2\ell+1$&$4\ell+5$&$4\ell-2^\dagger$&$4\ell-2$&$4\ell-1$&$3|n|+n+2\ell-2$\\
$\eta^{\ell-1}a_n C_{({\rm up})}^{\nu}(-2iz)^n\tilde{\Uhyp}_1 (m=0)$&$2|n|+2\ell-2$&$2|n|+2\ell+4$&$4\ell+8$&$4\ell+1^\dagger$&$4\ell+1$&$4\ell-1$&$3|n|+n+2\ell-2$\\
$\eta^{\ell-1}a_n C_{({\rm up})}^{\nu}(-2iz)^n\tilde{\Uhyp}_2 (m\neq 0) $&$4|n|-9$&$4|n|-3$&$4\ell+2$&$4\ell-2$&$4\ell-2^\dagger$&$4\ell+2$&$3|n|-n$\\
$\eta^{\ell-1}a_n C_{({\rm up})}^{\nu}(-2iz)^n\tilde{\Uhyp}_2 (m=0)$&$4|n|-6$&$4|n|$&$4\ell+5$&$4\ell+1$&$4\ell+1^\dagger$&$4\ell+2$&$3|n|-n$\\
\hline\hline
\end{tabular}
\caption{\label{tab:aU} The combined behaviour following from Tables~\ref{tab:a} and \ref{tab:U}. We take out a factor of $\eta^{l-1}$
for normalisation, fixing the largest term as $O(1)$; everything higher can be read as relative.}
 \end{center}
\end{table*}

\subsubsection{Phase extraction and the large \texorpdfstring{$\ell$}{l} modes: A PN ansatz}

As described in \cite{Kavanagh:2015a}, for sufficiently large values of $\ell$, in the
Schwarzschild case the MST solutions to the Regge-Wheeler equation, $X^{\text{in}/\text{up}}$ can
be written in the form:
\begin{align}
\XinMST&= e^{i \psi^\text{in}} X_1{}^{-\nu-1} \times
\nonumber \\
& \quad \left[1+\eta^2 A_2^{\ell}+\eta^4 A_4^{\ell}+\eta^6 A_6^{\ell}+\ldots\right] \\
\XupMST&= e^{i \psi^\text{up}} (X_2{}^{\!1/2}){}^{-\nu} \times
\nonumber \\
& \left[1+\eta^2 B_2^{\ell}+\eta^4 B_4^{\ell}+\dots+\eta^{2\ell} B_{2\ell}^{\ell}+O(\eta^{2\ell+2})\right] ,
\end{align}
where $\psi^{\text{in}}$ and $\psi^{\text{up}}$ are $r$-independent phase factors and the $A_i$'s
and $B_i$'s are pure polynomials in $X_1, \XtH$. (For low-$\ell$ values, this expression is
corrupted by logarithms and odd powers of $\eta$.)

At this point, two further optimisations help to dramatically improve the efficiency of
calculations. In constructing the retarded Green function, the phase factors will drop out since
they amount to an irrelevant normalisation. Throwing them away and working with the homogeneous
solutions without the phase factors leaves expressions which are orders of magnitude smaller in
complexity. A further simplification can be found by starting from the a large $\ell$ ansatz
\begin{align*}
A_{2n}^\ell=&  \sideset{}{'}\sum\limits_{i=0}^{n}  a_{(2n,i)} X_1{}^i  X_2{}^{\! n-i}, \\
 \nu(\epsilon) =& \, \ell+ \sum\limits_{j=1}^{\infty} a_{(6 j, 2 j)} \epsilon^{2 j}
\end{align*}
and solving for the $a_{(i,j)}$ by demanding we have a solution of the Regge-Wheeler equation. This
provides an efficient way to generate homogeneous solutions with $\ell,m$ left unspecified,
avoiding the complexity of the MST solutions.

We find that the situation with the Teukolsky equation is similar. Here the solutions can be re-expressed as
\begin{align}
{}_s \Rin&=e^{i \psi^\text{in}_{Kerr}} X_1{}^{-\nu-s} \times
\nonumber \\
& \qquad \left[1+\eta A_1^{\ell}+\eta^2 A_2^{\ell}+\eta^3 A_3^{\ell}+\eta^4 A_4^{\ell}+\ldots\right] \\
{}_s \Rup&=e^{i \psi^\text{up}_{Kerr}}(\XtH){}^{-\nu-1-s} \times
\nonumber \\
& \qquad \left[1+\eta B_1^{\ell}+\eta^2 B_2^{\ell}+\eta^3 B_3^{\ell}+\eta^4 B_4^{\ell}+\ldots\right] 
\end{align}
where once again up until an $\ell$-dependent power of $\eta$ the $A_i$ and $B_i$ are pure
polynomials in $X_1, \XtH$ and the essentially irrelevant phase functions remove significant
complexity. As in the Regge-Wheeler case, we proceed by using this as an ansatz for solutions with
$\ell,m$ unspecified and obtain general expressions for the large-$\ell$ homogeneous solutions.

\subsubsection{Teukolsky-Starobinsky identities}

In the previous section, the choice of spin, $s$, was arbitrary; $s=+2$ corresponds to $\psi_0$
while $s=-2$ corresponds to $\psi_4$. Only one or the other is required in order to obtain the full
radiative metric perturbation. However, in some situations it may be more convenient to have one or
the other. Fortunately, rather than repeating a lengthy calculation for both there is a convenient
shortcut. Given our set of spin $-2$ homogeneous solutions, we can easily calculate the spin $+2$
solution via a set of differential transformations known as the Teukolsky-Starobinsky identities.
These are discussed in detail in many places, for example see Ref.~\cite{Ori:2002uv}; for
completeness we repeat the final result here. Given either an ``in'' or an ``up'' solution
${}_{-2} \Rin/{}_{-2} \Rin$ of spin $s=-2$ one can write
\begin{align*}
{}_{+2} R^{\text{in}/\text{up}}_{\ell m \omega}=\frac{C^{\text{in}/\text{up}}}{\Delta^3}\left[A_0 \mathcal{D}_{\ell \omega}({}_{-2}R^{\text{in}/\text{up}}_{\ell m \omega})+B_0 {}_{-2}R^{\text{in}/\text{up}}_{\ell m \omega}\right]
\end{align*}
where
\begin{align*}
\mathcal{D}_{\ell \omega} = & \partial_r- \frac{i K}{\Delta} \\
A_0 = & 8 i K\big[K^2+(r-M)^2\big] \\
& -\big[4 i K (\lambda_{CH}+2)-8 i \omega r(r-M)\big]\Delta+8 i \omega \Delta^2 \\
B_0=&\big[(\lambda_{CH}+2-2i \omega r)(\lambda_{CH}+6 i \omega r) \\
  & \quad -12 i \omega (i K+r-M)\big] \Delta \\
  & + 4 i K [i K+r-M][\lambda_{CH}+6 i \omega r] \\
\lambda_{CH}&=\lambda_{MST}+s+|s|.
\end{align*}
It is worth noting that on the computational side, upon doing this transformation, we will reintroduce $r$-independent terms of the form $X_1 \XtH$ that can once again be extracted as components in the phase and essentially ignored. 

\subsubsection{Spheroidal functions}
\label{sec:spheroidal}

For the purposes of our calculation the only relevant frequencies are multiples of the orbital
frequency, which in the PN regime is asypmtotically small. For the calculation of the spin-weighted
spheroidal functions this allows us to use a low frequency perturbative expansion in terms of the
spin- weighted spherical harmonics. Recall the ODE for the spin-weighted spheroidal harmonics,
Eq.~(\ref{Eq:Slmeq}), which can be rewritten suggestively as
\begin{align}
(\mathcal{L}^0+a\omega \mathcal{L}^1){}_s\Slm(\theta,\varphi;a \omega)= - {}_s \eig \, {}_s\Slm(\theta,\varphi;a \omega)  \label{eq:Sphperturb}
\end{align}
where 
\begin{align}
\mathcal{L}^0&=\frac{1}{\sin \theta}\diff{}{\theta}\left(\sin\theta \diff{  }{\theta}\right)
-\frac{(m+s\cos\theta)^2}{\sin^2\theta} + s \\
\mathcal{L}^1&=-2s\cos\theta+2m-a\omega(1-\cos^2\theta).
\end{align}
Then, Eq.~(\ref{eq:Sphperturb}) reduces to an ODE for the spin-weighted spherical harmonics in the $a\omega\rightarrow 0$ limit.
This suggests a perturbative solution
\begin{equation}
\vphantom{}_s S_{lm} (\theta, \varphi;a \omega)=\sum_{k=|s|}^\infty d_{km}(a\omega) \vphantom{}_s Y_{km} (\theta, \varphi),
\end{equation}
where the series coefficients are written as expansions in $a\omega$,
\begin{align}
d_{km}&=1+d_{km}^1(a \omega)+d_{km}^2(a\omega)^2+ \cdots.
\end{align}
Similarly, we may write the eigenvalue as a series in $a\omega$,
\begin{equation}
{}_s\eig=\ell(\ell+1)-s(s+1)+\lambda^{(1)}(a \omega)+\lambda^{(2)}(a \omega)^2+\cdots.
\end{equation}
Then, using the relations
\begin{align}
\mathcal{L}^0{}_s Y_{\ell m} & (\theta, \varphi)= - [\ell(\ell+1)-s(s+1)]{}_s Y_{\ell m}(\theta, \varphi), \nonumber \\
\cos\theta{}_s Y_{\ell m} & (\theta, \varphi)={}_s \alpha_{\ell m} \, {}_sY_{\ell-1,m }(\theta, \varphi) \nonumber \\
  & +{}_s \beta_{\ell m} \, {}_sY_{\ell m}(\theta, \varphi)+{}_s \alpha_{\ell+1,m} \, {}_sY_{\ell+1,m}(\theta, \varphi), \nonumber \\
\cos^2 \theta{}_s Y_{\ell m} & (\theta, \varphi)={}_s \alpha_{\ell-1, m} \, {}_s \alpha_{\ell, m} \, {}_sY_{\ell-2,m }(\theta, \varphi) \nonumber \\
 & + {}_s \alpha_{\ell m} \, ({}_s \beta_{\ell-1, m} \, + {}_s \beta_{\ell m} ){}_sY_{\ell-1,m }(\theta, \varphi) \nonumber \\
  & +({}_s \alpha_{\ell m}^2+{}_s \alpha_{\ell+1, m}^2+{}_s \beta_{\ell m}^2) \, {}_sY_{\ell m}(\theta, \varphi) \nonumber \\
 & + {}_s \alpha_{\ell+1, m} \, ({}_s \beta_{\ell m} \, + {}_s \beta_{\ell+1, m} ){}_sY_{\ell+1,m }(\theta, \varphi) \nonumber \\
  & + {}_s \alpha_{\ell+1,m} \, {}_s \alpha_{\ell+2,m} \, {}_sY_{\ell+2,m}(\theta, \varphi),
\end{align}
with
\begin{align}
{}_s \alpha_{\ell m} &= \frac{\sqrt{\left(\ell^2-m^2\right) \left(\ell^2-s^2\right)}}{\ell \sqrt{(2 \ell+1) (2\ell-1)}},	\\
{}_s \beta_{\ell m} &= -\frac{m s}{\ell(\ell+1)},
\end{align}
we can use the orthogonality of the spherical harmonics to reduce the problem to solving a system
of linear algebraic equations for the series coefficients at each $a\omega$ order. This process
determines the coefficients $\lambda^{i}$ and $d^i_{km}$ for $k\ne 0$; the remaining coefficients,
$d^i_{0m}$, are then determined by enforcing normalisation of the spin-weighted spheroidal harmonics
via Eq.~\eqref{eq:spheroidal-normalisation}. In Appendix \ref{sec:spheroidal-expansion} we give an
explicit expression for this expansion for general spin, $s$, to order $(a\omega)^4$.

\subsection{Reconstructed metric perturbation in radiation gauge}
\label{sec:metric-reconstruction}

The procedure for building up the components of the metric perturbation from 
the Weyl scalars \cite{Chrzanowski:1975wv,Cohen:1974cm,Kegeles:1979an,Wald:1978vm,Stewart:1978tm} 
involves the construction of a Hertz potential, $\Psi$, from which gives the metric perturbation can
be computed using
\begin{align}
h_{\alpha\beta}&= - \varrho^{-4}\{n_\alpha n_\beta(\boldsymbol{\bar{\delta}}-3\alpha-\bar{\beta}+5\varpi)(\boldsymbol{\bar{\delta}}-4\alpha+\varpi)\nonumber\\
&+\bar{m}_\alpha \bar{m}_\beta(\boldsymbol{\Delta}+5\mu-3\gamma-\bar{\gamma})(\boldsymbol{\Delta}+\mu-4\gamma)\nonumber\\
&-n_{(\alpha}n_{\beta)}\left[(\boldsymbol{\bar{\delta}}-3\alpha+\bar{\beta}+5\varpi+\bar{\tau})(\boldsymbol{\Delta}+\mu-4\gamma)\right. \nonumber \\
&\left. +(\boldsymbol{\Delta}+5 \mu -\bar{\mu}-3\gamma -\bar{\gamma})(\boldsymbol{\bar{\delta}}-4\alpha+\varpi)\right]\}\Psi+\text{c.c.}, \label{Eq:hofPsi}
\end{align}
where the overall minus sign here accounts for our mostly positive sign convention for the metric.

For simplicity, from this point forward we restricting ourselves to the outgoing radiation gauge
(ORG) defined by
\begin{align*}
n^{\mu}h_{\mu\nu}=0.
\end{align*}
In this gauge the Hertz potential itself is a solution of the spin-$2$ Teukolsky equation.
Calculating $\Psi$ can be done either using $\psi_0$ or $\psi_4$ in a variety of ways. For example,
in \cite{vandeMeent:2015lxa} expressions are given for the $\Psi$ in terms of the asymptotic
amplitudes of $\psi_4$, whereas in \cite{Keidl:2010pm} it is constructed by inverting a
differential operator which simplifies in the circular orbit case. Therefore to proceed one must
solve either the $s=2$ or $s=-2$ Teukolsky equation. However from a practical point of view it is
not particularly important which is chosen, as the Teukolsky-Starobinsky identities can be used 
to transform between the two.

In this work we chose to construct
$\psi_0$ using the $s=2$ homogeneous solutions. To do this we construct the retarded Green function
\begin{align}
G_{\ell m}(r,r')=-\frac{{}_2 R^{\text{in}}_{\ell m}(r_<) {}_2 R^{\text{up}}_{\ell m}(r_<)}{A_{\ell m}},
\end{align}
where $A_{\ell m}=\Delta^{s+1}W(R^{\text{in}},R^{\text{up}})$ is the invariant Wronskian. Then 
\begin{align}
\psi_{0,\ell m\omega}=4 \pi\int \Delta'^2 G_{\ell m}(r,r') & {}_2 S^\ast_{\ell m}(\theta',\varphi';a \omega)T_0' \nonumber \\
&\times  \Sigma' \sin\theta' dr' d\theta' d\varphi'
\end{align}
Then, using Eq.~(40) of \cite{Keidl:2010pm}
\begin{align}
&\Psi_{\ell m}= \nonumber \\
& \quad 8\frac{(-1)^m D \bar{\psi}_{0,\ell,-m,-\omega}+12 i M \omega \psi_{0,\ell m\omega} }{D^2+144 M^2 \omega^2} {}_2S_{\ell m}(\theta, \varphi; a \omega),
\end{align}
where 
\begin{align*}
D^2=&\lambda_{CH}^2(\lambda_{CH}+2)^2+8 a \omega(m-a \omega)\lambda_{CH}(5 \lambda_{CH}+6) \nonumber \\
&+48 (a\omega)^2(2\lambda_{CH}+3(m-a\omega)^2)
\end{align*}
and $\lambda_{CH}=\lambda_{MST}+s+|s|$.

Finally, we construct the $\ell,m$ modes of the metric components, $h_{\alpha\beta}^{\ell m}$, using 
\eqref{Eq:hofPsi} with $\Psi$ replaced by $\Psi_{\ell m}$. Note that this is is in contrast to a mode definition in terms of 
a direct decomposition of $h_{\alpha\beta}$ over the spin-weighted spheroidal harmonics; the modes of the metric
perturbation constructed in this way are not necessarily pure spin $\pm2$ spheroidal harmonic modes.

\subsection{Metric completion}
\label{sec:completion}
It is well known that a metric reconstruction procedure based on $\psi_0$ or $\psi_4$ does not
yield the whole radiation-gauge metric perturbation. Wald \cite{Wald:1973} showed that for a Kerr
background the remaining part of the metric perturbation can be fully attributed to perturbations
to the mass and angular momentum of the background black hole (often, these are informally
described as ``$\ell=0$ and $\ell=1$ parts'' of the perturbation). Here, we follow the standard
procedure \cite{Shah:2012gu} and incorporate this contribution using analytic expressions. In
particular, defining $H = \tfrac12 h_{\alpha \beta} u^\alpha u^\beta$, the additional contributions
corresponding to perturbations to the mass and angular momentum are
\begin{align}
  H_{\delta M} =&
\frac{\left(r_0^2+2 a \sqrt{M r_0}-a^2\right)\left(r_0^{3/2}-2 M \sqrt{r_0}+a \sqrt{M}\right)
}{r_0^{9/4} \left(r_0^{3/2}-3 M\sqrt{r_0}+2 a \sqrt{M}\right)^{3/2}}, \\ H_{\delta J} =&
\frac{\sqrt{M} \left(r_0^2-2 a \sqrt{M r_0}+a^2\right) \left(a-2 \sqrt{M r_0}\right) }{r_0^{9/4}
\left(r_0^{3/2}-3 M\sqrt{r_0}+2 a \sqrt{M}\right)^{3/2}}.
\end{align}
There is a subtlety here in that these contributions are not smooth on the worldline, and this
non-smoothness could introduce additional contributions to the regularization procedure (see
\cite{Pound:2013faa} for a more detailed discussion). This is a complicated issue worthy of a
detailed independent analysis; here, we merely follow the standard procedure of evaluating the
contributions from the completion part in the limit $r \to r_0^+$ and use the mode-sum
regularization procedure described in the next section.

\subsection{Regularization}
We adopt a variation of the standard mode-sum regularization approach in order
to extract a finite value from the divergent retarded metric perturbation.
Traditionally, this mode-sum approach is written in terms of a sum over regularized
\emph{spherical} harmonic modes. This has the distinct disadvantage of requiring
a cumbersome projection of the modes of the retarded metric perturbation onto scalar
spherical harmonics. We have avoided this unnecessary step by instead deriving a mode-sum
formula for the \emph{spheroidal} harmonic modes that naturally arise from solutions of
the Teukolsky equation. The full details of this derivation will be given in a forthcoming
work; here we merely highlight the key results.

The derivation of our mode-sum formula is conceptually similar to previous
derivations in terms of spherical harmonics, i.e.
\begin{enumerate}
\item Work in a spherical coordinate system, $(\alpha, \beta)$ in which the particle
      is instantaneously located at the north pole, $\alpha=0$.
\item Obtain a local coordinate expansion of the contribution to $H$ from the 
      Detweiler-Whiting singular field.
\item Decompose this coordinate approximation into spin-0 spheroidal harmonics,
      where the spheroidal harmonics are defined with respect to a coordinate system
      $(\theta, \varphi)$ in which the worldline is in the equatorial plane, $\theta = 0$.
      The decomposition process makes use of the relation between the modes in the 
      $(\alpha, \beta)$ coordinate system (where the mode decomposition can be
      most-easily done analytically) and the modes in the $(\theta, \varphi)$ coordinate
      system (where the retarded-field modes are most easily obtained).
\item Sum over $m$ (azimuthal) modes to obtain a mode-sum formula.
\end{enumerate}
The result of this process is a mode-sum formula for computing the regularized
redshift invariant,
\begin{equation}
  H^R = \sum_{\ell=0}^\infty \big(H^{\rm ret}_{\ell} - H_{[0]}\big),
\end{equation}
where
\begin{align}
  H_{[0]} &= \frac{2}{\pi \zeta} \mathcal{K}
  + \frac{a^2 \Omega^2}{6 \pi \zeta k^2} \Big[(k-2) \mathcal{E}-2(k-1) \mathcal{K}\Big] \nonumber \\ & \quad
  + \frac{a^4 \Omega^4}{1120 \pi  \zeta  k^4} \Big[
    2 (9 k^3 + 4 k^2+116 k-152) \mathcal{E} \nonumber \\ & \qquad
    - (9 k^3-89 k^2+384k-304) \mathcal{K}
  \Big] + \mathcal{O}\Big(\frac{a^6}{r_0^{10}}\Big)
\end{align}
with
\begin{equation*}
  \zeta^2 \equiv \mathcal{L}^2 + r_0^2 + \frac{2 a^2 M}{r_0} + a^2, \quad
  k\equiv\frac{\zeta^2 - r_0^2}{\zeta^2},
\end{equation*}
and with $\mathcal{K}(k)$ and $\mathcal{E}(k)$ being complete elliptic integrals.
For the purposes of this work, we require
the post-Newtonian expansion of this regularization parameter, which is given by \\
\begin{widetext}
\begin{align}
H_{[0]} &=
  y
  - \frac{y^2}{4}
  + \frac{2}{3} q y^{5/2}
  - \left(\frac{39}{64}+\frac{q^2}{4}\right) y^3
  + \frac{7}{6} q y^{7/2}
  - \left(\frac{385}{256}+\frac{7q^2}{36}\right) y^4
  + \left(\frac{99 q}{32}-\frac{q^3}{2}\right) y^{9/2} \nonumber \\ & \quad
  - \left(\frac{61559}{16384}+\frac{1625 q^2}{2304}-\frac{9 q^4}{64}\right) y^5
  + \left(\frac{3239 q}{384}-\frac{733 q^3}{648}\right) y^{11/2}
  - \left(\frac{622545}{65536}+\frac{5827q^2}{2048}+\frac{41 q^4}{256}\right) y^6 \nonumber \\ & \quad
  + \left(\frac{577769 q}{24576}-\frac{27133 q^3}{10368}+\frac{15 q^5}{32}\right) y^{13/2}
  - \left(\frac{25472511}{1048576}+\frac{6885521 q^2}{589824}+\frac{178879 q^4}{248832}+\frac{25 q^6}{256}\right) y^7 \nonumber \\ & \quad
  + \left(\frac{2183421q}{32768}-\frac{2173 q^3}{512}+\frac{173 q^5}{128}\right) y^{15/2}
  - \left(\frac{263402721}{4194304}+\frac{54148187 q^2}{1179648}+\frac{6253225 q^4}{1990656}-\frac{861 q^6}{2048}\right) y^8 \nonumber \\ & \quad
  + \left(\frac{100247739 q}{524288}+\frac{2495743 q^3}{2654208}+\frac{3176285 q^5}{746496}-\frac{175q^7}{384}\right) y^{17/2} \nonumber \\ & \quad
  -\left(\frac{176103411255}{1073741824}+\frac{722675577 q^2}{4194304}+\frac{13050523 q^4}{1048576}-\frac{18421 q^6}{8192}-\frac{1225 q^8}{16384}\right) y^9 \nonumber \\ & \quad
  + \left(\frac{1161008301 q}{2097152}+\frac{544444555 q^3}{10616832}+\frac{34364327 q^5}{2985984}-\frac{6335 q^7}{4608}\right) y^{19/2} + \mathcal{O}(y^{10}).
\end{align}

\section{Results}
\label{sec:results}
The main result of this work is the post-Newtonian expansion of Detweiler's redshift invariant. This
is given as a series expansion in $y$ and $\log y$, which takes the form
\begin{align}
  \Delta U &= 
      c_1 y + c_2 y^2 + c_{2.5} y^{2.5} + c_3 y^3 + c_{3.5} y^{3.5} + c_{4} y^{4} 
      + c_{4.5} y^{4.5} + (c_{5} + c^{\rm ln}_{5} \log y) y^{5} + c_{5.5} y^{5.5}
      + (c_{6} + c^{\rm ln}_{6} \log y) y^{6}  \nonumber \\
  & \quad
      + (c_{6.5} + c_{6.5}^{\rm ln} \log y) y^{6.5}  
      + (c_{7} + c^{\rm ln}_{7} \log y) y^{7} + (c_{7.5} + c_{7.5}^{\rm ln} \log y)  y^{7.5}
      + (c_{8} + c^{\rm ln}_{8} \log y + c^{\rm ln^2}_{8} \log^2 y) y^{8}  \nonumber \\
  & \quad 
      + (c_{8.5} + c_{8.5}^{\rm ln} \log y)  y^{8.5} 
      + (c_{9} + c^{\rm ln}_{9} \log y + c^{\rm ln^2}_{9} \log^2 y) y^{9}
      + (c_{9.5} + c_{9.5}^{\rm ln} \log y + c^{\rm ln^2}_{9.5} \log^2 y)  y^{9.5}
      + \mathcal{O}(y^{10}).
\end{align}
The coefficients in this expansion are given by
\begin{gather*}
c_{1} = - 1, \quad
c_{2} = -2, \quad
c_{2.5} = \tfrac{7}{3} q, \quad
c_{3} = -5 -  q^2, \quad
c_{3.5} = \tfrac{46}{3} q, \quad
c_{4} = - \tfrac{121}{3} + \tfrac{41}{32} \pi^2 -  \tfrac{86}{9} q^2, \quad
c_{4.5} = 77 q + q^3, \nonumber \\ 
c_{5} = - \tfrac{1157}{15} -  \tfrac{128}{5} \gamma + \tfrac{677}{512} \pi^2 -  \tfrac{256}{5} \log(2) -  \tfrac{577}{9} q^2, \quad
c_{5}^{\rm ln} = - \tfrac{64}{5}, \quad
c_{5.5} = \big[\tfrac{974}{3} + \tfrac{29}{32} \pi^2\big] q + \tfrac{1526}{81} q^3,
\end{gather*}
\begin{align}
c_{6} &= \tfrac{1606877}{3150} + \tfrac{1912}{105} \gamma -  \tfrac{60343}{768} \pi^2 + \tfrac{7544}{105} \log(2) -  \tfrac{243}{7} \log(3) - \big[\tfrac{1147}{3} - \tfrac{593}{512} \pi^2 \big] q^2 - 2 q^4, \nonumber \\ 
c_{6}^{\rm ln} &= \tfrac{956}{105}, \nonumber \\ 
c_{6.5} &= - \tfrac{13696}{525} \pi + q \big[\tfrac{348047}{150} + \tfrac{352}{5} \gamma -  \tfrac{6349}{64} \pi^2 + \tfrac{416}{3} \log(2)\big] + \tfrac{13625}{81} q^3, \nonumber \\ 
c_{6.5}^{\rm ln} &= \tfrac{176}{5} q, \nonumber \\ 
c_{7} &= \tfrac{17083661}{4050} + \tfrac{102512}{567} \gamma -  \tfrac{1246056911}{1769472} \pi^2 + \tfrac{2800873}{262144} \pi^4 + \tfrac{372784}{2835} \log(2) + \tfrac{1215}{7} \log(3) \nonumber \\ 
& \quad - q^2 \big[ \tfrac{1288408}{675} +  \tfrac{264}{5} \gamma - \tfrac{92557}{9216} \pi^2 + 104 \log(2)\big] -  \tfrac{8120}{243} q^4 , \nonumber \\ 
c_{7}^{\rm ln} &= \tfrac{51256}{567} -  \tfrac{132}{5} q^2, \nonumber \\ 
c_{7.5} &= \tfrac{81077}{3675} \pi + q \big[\tfrac{734961481}{22050} + \tfrac{2072}{5} \gamma -  \tfrac{8911441}{3072} \pi^2 + \tfrac{4744}{7} \log(2) + \tfrac{972}{7} \log(3) + \tfrac{32}{5} \log(\kappa) + \tfrac{16}{5} \psi^{(0,2)}(q)\big] \nonumber \\ 
& \quad + q^3 \big[\tfrac{243611}{225} + \tfrac{96}{5} \gamma + \tfrac{1319}{384} \pi^2 + \tfrac{96}{5} \log(2) + \tfrac{96}{5} \log(\kappa) + \tfrac{48}{5} \psi^{(0,2)}(q)\big] + \tfrac{12}{5} q^5, \nonumber \\ 
c_{7.5}^{\rm ln} &= \tfrac{1052}{5} q + \tfrac{96}{5} q^3, \nonumber \\ 
c_{8} &= \tfrac{12624956532163}{382016250} -  \tfrac{10327445038}{5457375} \gamma + \tfrac{109568}{525} \gamma^2 -  \tfrac{9041721471697}{2477260800} \pi^2 -  \tfrac{23851025}{16777216} \pi^4 -  \tfrac{16983588526}{5457375} \log(2) \nonumber \\ 
& \quad + \tfrac{438272}{525} \gamma \log(2) + \tfrac{438272}{525} \log(2)^2 - \tfrac{2873961}{24640} \log(3) -  \tfrac{1953125}{19008} \log(5) -  \tfrac{2048}{5} \zeta(3)\nonumber \\ 
& \quad + \tfrac{33008}{315} \pi q - q^2 \big[-\tfrac{14713942}{945} + \tfrac{67736}{315} \gamma + \tfrac{710125279}{294912} \pi^2 +  \tfrac{6632}{21} \log(2) + \tfrac{729}{7} \log(3)\big] - \big[ \tfrac{85420}{243} + \tfrac{69}{256} \pi^2\big] q^4 , \nonumber \\ 
c_{8}^{\rm ln} &= - \tfrac{5163722519}{5457375} + \tfrac{109568}{525} \gamma -  \tfrac{33868}{315} q^2 + \tfrac{219136}{525} \log(2), \nonumber \\ 
c_{8}^{\rm ln^{2}} &= \tfrac{27392}{525}, \nonumber \\ 
c_{8.5} &= \tfrac{82561159}{467775} \pi + q \big[\tfrac{700704798839}{3572100} + \tfrac{1097696}{8505} \gamma -  \tfrac{27925459441}{1327104} \pi^2 + \tfrac{124925059}{393216} \pi^4 + \tfrac{4440032}{8505} \log(2) -  \tfrac{162}{7} \log(3) \nonumber \\ 
& \quad + \tfrac{224}{5} \log(\kappa) + \tfrac{16}{5} \psi^{(0,1)}(q) + \tfrac{96}{5} \psi^{(0,2)}(q)\big] -  \tfrac{5564}{105} \pi q^2 + q^3 \big[\tfrac{13856317}{1215} + \tfrac{2792}{15} \gamma -  \tfrac{80954347}{165888} \pi^2 + \tfrac{1272}{5} \log(2) \nonumber \\ 
& \quad + \tfrac{552}{5} \log(\kappa) -  \tfrac{12}{5} \psi^{(0,1)}(q) + \tfrac{288}{5} \psi^{(0,2)}(q)\big] + \tfrac{191699}{3645} q^5 , \nonumber \\ 
c_{8.5}^{\rm ln} &= \tfrac{147872}{1701} q + \tfrac{2224}{15} q^3, \nonumber \\ 
c_{9} &= - \tfrac{7516468355368067}{34763478750} -  \tfrac{1533327047906}{496621125} \gamma -  \tfrac{108064}{2205} \gamma^2 -  \tfrac{246847155756529}{18496880640} \pi^2 + \tfrac{22759807747673}{6442450944} \pi^4 + \tfrac{1712}{525} \kappa \nonumber \\
& \quad -  \tfrac{1363551923554}{496621125} \log(2) -  \tfrac{3574208}{3675} \gamma \log(2) -  \tfrac{2143328}{1575} \log(2)^2 -  \tfrac{2201898578589}{392392000} \log(3) + \tfrac{37908}{49} \gamma \log(3) \nonumber \\
& \quad + \tfrac{37908}{49} \log(2) \log(3) + \tfrac{18954}{49} \log(3)^2 + \tfrac{798828125}{741312} \log(5) -  \tfrac{64}{5} \log(\kappa) -  \tfrac{32}{5} \psi^{(0,2)}(q) -  \tfrac{41408}{105} \zeta(3) \nonumber \\
& \quad + q \big[\tfrac{24020077}{66150} \pi -  \tfrac{3424}{525} \bar{\psi}^{(0,2)}(q) + \tfrac{64}{5 \kappa}  \bar{\psi}^{(1,2)}(q)\big] + q^2 \big[\tfrac{53568869587}{99225} -  \tfrac{682000}{567} \gamma -  \tfrac{411304830035}{7077888} \pi^2 -  \tfrac{417436343}{16777216} \pi^4 + \tfrac{1712}{175} \kappa \nonumber \\
& \quad -  \tfrac{7102544}{2835} \log(2) + \tfrac{486}{7} \log(3) -  \tfrac{448}{3} \log(\kappa) -  \tfrac{64}{15} \psi^{(0,1)}(q) -  \tfrac{352}{5} \psi^{(0,2)}(q)\big] - q^3 \big[\tfrac{3424}{175} \bar{\psi}^{(0,2)}(q) - \tfrac{192}{5 \kappa} \bar{\psi}^{(1,2)}(q)\big] \nonumber \\
& \quad - q^4 \big[\tfrac{650593}{225} + \tfrac{104}{5} \gamma - \tfrac{818819}{98304} \pi^2 + \tfrac{232}{5} \log(2) - \tfrac{32}{5} \log(\kappa) - \tfrac{16}{5} \psi^{(0,1)}(q)\big] -  \tfrac{14}{5} q^6, \nonumber \\
c_{9}^{\rm ln} &= - \tfrac{769841899153}{496621125} -  \tfrac{108064}{2205} \gamma -  \tfrac{1787104}{3675} \log(2) + \tfrac{18954}{49} \log(3) -  \tfrac{383336}{567} q^2 -  \tfrac{36}{5} q^4, \nonumber \\
c_{9}^{\rm ln^{2}} &= - \tfrac{27016}{2205}, \nonumber \\
c_{9.5} &= - \tfrac{2207224641326123}{1048863816000} \pi + \tfrac{23447552}{55125} \gamma \pi -  \tfrac{219136}{1575} \pi^3 -  \tfrac{10755481}{33075} \pi q^2 + \tfrac{46895104}{55125} \pi \log(2) \nonumber \\
& \quad + q \big[\tfrac{2167536532386661}{2521307250} + \tfrac{75699353672}{16372125} \gamma -  \tfrac{437824}{525} \gamma^2 -  \tfrac{780002666754601}{7431782400} \pi^2 + \tfrac{43593199495}{16777216} \pi^4 + \tfrac{144895599176}{16372125} \log(2) \nonumber \\
& \quad -  \tfrac{5240192}{1575} \gamma \log(2) -  \tfrac{1744448}{525} \log(2)^2 -  \tfrac{10841769}{6160} \log(3) + \tfrac{9765625}{14256} \log(5) + \tfrac{10208}{35} \log(\kappa) + \tfrac{2806}{105} \psi^{(0,1)}(q) \nonumber \\
& \quad + \tfrac{12416}{105} \psi^{(0,2)}(q) + \tfrac{6}{7} \psi^{(0,3)}(q) + \tfrac{5504}{5} \zeta(3)\big]
+ q^3 \big[\tfrac{164687954986}{297675} + \tfrac{5178416}{2835} \gamma -  \tfrac{17476082953}{331776} \pi^2 + \tfrac{7481392}{2835} \log(2) \nonumber \\
& \quad + \tfrac{1944}{7} \log(3) + \tfrac{71984}{105} \log(\kappa) -  \tfrac{1307}{70} \psi^{(0,1)}(q) + \tfrac{7424}{21} \psi^{(0,2)}(q) + \tfrac{111}{14} \psi^{(0,3)}(q)\big] + q^5 \big[\tfrac{1079765333}{1786050} + \tfrac{272}{35} \gamma -  \tfrac{53}{1024} \pi^2 \nonumber \\
& \quad + \tfrac{272}{35} \log(2) + \tfrac{272}{35} \log(\kappa) -  \tfrac{36}{35} \psi^{(0,1)}(q) -  \tfrac{128}{35} \psi^{(0,2)}(q) + \tfrac{60}{7} \psi^{(0,3)}(q)\big], \nonumber \\
c_{9.5}^{\rm ln} &= \tfrac{11723776}{55125} \pi + q \big[\tfrac{40237200436}{16372125} -  \tfrac{437824}{525} \gamma -  \tfrac{2620096}{1575} \log(2)\big] + \tfrac{3560992}{2835} q^3 + \tfrac{272}{35} q^5, \nonumber \\
c_{9.5}^{\rm ln^{2}} &= - \tfrac{109456}{525} q
\end{align}
where $\gamma$ is Euler's constant, $\zeta(n)$ is the Riemann zeta function, $\psi^{(n,k)} (q) \equiv \psi^{(n)}(\tfrac{i k q}{\kappa}) + \psi^{(n)}(\tfrac{-i k q}{\kappa}) = 2 \Re[\psi^{(n)}(\tfrac{i k q}{\kappa})]$, $\bar{\psi}^{(n,k)} (q) \equiv - i [\psi^{(n)}(\tfrac{i k q}{\kappa}) - \psi^{(n)}(\tfrac{-i k q}{\kappa})] = 2 \Im[\psi^{(n)}(\tfrac{i k q}{\kappa})]$
and $\psi^{(n)}(z) = \frac{d^{n+1}}{dz^{n+1}}\ln \Gamma(z)$ is the polygamma function. For
convenience, we have also made these expressions available online as \textsc{Mathematica} notebooks
\cite{online}.
\end{widetext}

\section{Discussion}
\label{sec:discussion}

In this work, we have presented results for the linear-in-mass ratio contribution to Detweiler's
redshift invariant, $\Delta U$, for a quasi-circular binary black hole system, in the case where the
larger black hole is spinning. Our results are given as a PN-type expansion in the inverse
separation, $y$, of the binary, but are otherwise exact. In particular, they are valid for
astrophysically-important cases where the spin of the larger black hole is arbitrarily large.

There are several clear future directions for this work, in particular:
\begin{itemize}
\item An extension to eccentric and inclined orbits would allow for a more complete exploration of
      the parameter space. The components for such a calculation are readily available --- eccentric
      orbits have been studied without spin in \cite{Bini:2015bfb,Hopper:2015icj,Bini:2016qtx},
      while spin contributions have been considered here and in \cite{Bini:2015xua} --- so such a
      calculation would merely require an appropriate combination of the two approaches.
\item An extension to higher PN orders. As the method we present here is totally algorithmic and
      implemented as a \textsc{Mathematica} code, it is straightforward (but more computationally
      expensive) to apply it to higher orders if the demand arises. Indeed, an online repository of
      our results \cite{online} will be updated as future results become available.
\item An extension to second order in the mass ratio would enable us to probe potentially important
      non-linear effects. This would require substantial effort, but recent progress
      \cite{Detweiler:2011tt,Pound:2012nt,Gralla:2012db,Pound:2012dk,Pound:2014xva,Pound:2014koa,Pound:2015fma,Pound:2015tma,Pound:2015wva}
      indicates rapid progress towards this goal.
\item The application of our method to the calculation of other gauge invariant quantities, such as
      the spin-precession \cite{Dolan:2013roa}, tidal \cite{Dolan:2014pja}, and octupolar
      \cite{Nolan:2015vpa} invariants. These are obtained from the same metric perturbations used
      for the redshift invariant, so their calculation would be a straightforward application of the
      results presented here.
\end{itemize}
In addition to these, it is likely that an application of our results to improving and informing
PN and EOB theories --- for example by computing GSF contributions to the potentials appearing in EOB theory --- would yield valuable improvements to both PN and EOB models.

\acknowledgments
We thank Abhay Shah for performing numerical checks of our results. We also thank Abhay Shah,
Bernard Whiting, Seth Hopper, Niels Warburton, Marc Casals and Sarp Akcay for many helpful discussions.
B.W. was supported by the Irish Research Council, which is
funded under the National Development Plan for Ireland. C.K. was
funded under the Programme for Research in Third Level
Institutions (PRTLI) Cycle 5 and co-funded under the European
Regional Development Fund. This material is based upon work supported
by the National Science Foundation under Grant Number 1417132.

\appendix

\section{Sums over \texorpdfstring{$m$}{m}-modes}
Using the methods of the previous sections, we can write the $\ell,m$-modes of the metric
perturbation in terms of radial functions multiplying combinations of the spin-weighted spheroidal
harmonics and their $\theta$ derivatives. When limiting to the position of the particle and summing
over $m$, one then encounters sums such as
\begin{align}
S_1^{N}&=\sum_{m=-\ell}^{\ell}m^N |{}_s\Slm(\pi/2,0; a m \Omega)|^2 \label{eq:samplesumes1}\\
S_2^{N}&=\sum_{m=-\ell}^{\ell}m^N {}_s\Slm(\pi/2,0; a m \Omega) \partial_\theta \times \nonumber \\
& \qquad \qquad {}_s \Slm^\ast(\pi/2,0; i a m \Omega) \label{eq:samplesumes2}\\
S_3^{N}&=\sum_{m=-\ell}^{\ell}m^N |\partial_\theta {}_s \Slm(\pi/2,0; a m \Omega)|^2  \label{eq:samplesumes3}
\end{align}
It is not immediately clear how to do these explicitly, but for the low-frequency limit we are
interested in, progress can be made by using an expansion in terms of spin-weighted spherical
harmonics, and doing the sums order by order. As an example, for $S_1^N$ with $s=2$ we find
\begin{widetext}
\begin{align}
S_1^N&=\sum_{m=-\ell}^{\ell} \left\{m^N  {}_2\Ylm(\pi/2,0)^2 \nonumber +4 m^{N+1} q \Omega \left[\frac{\sqrt{(\ell-2)(\ell+2)(\ell+m)(\ell-m)}}{\ell^2\sqrt{(2\ell-1)(2\ell+1)}}\, {}_2Y_{\ell-1,m}(\pi/2,0)\, {}_2Y_{\ell m}(\pi/2,0) \right. \right. \nonumber \\
&\qquad \qquad \qquad \left.\left.-\frac{\sqrt{(\ell-1)(\ell+3)(\ell+1+m)(\ell+1-m)}}{(\ell+1)^2\sqrt{(2\ell+1)(2\ell+3)}}\, {}_2Y_{\ell+1,m}(\pi/2,0)\, {}_2Y_{\ell m}(\pi/2,0) \right] + O\big[(q \Omega)^2\big]\right\}
\end{align}
which is zero for odd values of $N$. For even values we find it efficient to obtain closed form
expressions each sum using Mathematica's \texttt{FindSequenceFunction} routine. This function takes
as an argument a sample range of algebraic evaluations of one of these sums for given values of
$\ell$ and outputs an analytic form for general $\ell$. The method is highly parallelisable, which
given the number of different variations of sums we will face for increasing order, and
the different combinations of ${}_s \Slm(\pi/2,0; a m \Omega)$ and its derivative, is
extremely useful. Example results for $N=0,2$ are
\begin{align*}
S_1^0 &=\frac{2\ell+1}{4 \pi}\bigg[1-\frac{16 }{  \ell^2 (\ell+1)^2} a \Omega-  \frac{1}{ 2 \ell^4 (\ell+1)^4 (2 \ell-1)^2 (2 \ell+3)^2} \times \nonumber \\
& \qquad (-14688+9792 \ell+49560 \ell^2+60528 \ell^3-15352
   \ell^4-49411 \ell^5-7597 \ell^6+7598 \ell^7+1904 \ell^8 \nonumber \\
& \qquad \qquad +25 \ell^9+27 \ell^{10}+12 \ell^{11}+2 \ell^{12}) a^2 \Omega^2
\bigg]+ O\big[(q \Omega)^3\big]\\
S_1^2 &=\frac{2\ell+1}{4 \pi}\bigg[\frac{1}{2} \left(-4+\ell+\ell^2\right)-\frac{\left(-68+3 \ell^2+6 \ell^3+3 \ell^4\right) }{\ell^2
   (\ell+1)^2}a\Omega-\frac{1}{8 \ell^4 (\ell+1)^4 (2\ell-1)^2 (2\ell+3)^2} \times \nonumber\\
& \qquad (428544-285696 \ell-1487712 \ell^2-1738176 \ell^3+707612 \ell^4+1645334
   \ell^5+136153 \ell^6\nonumber \\
& \qquad \qquad -354650 \ell^7-88019 \ell^8+3938 \ell^9+4411 \ell^{10}+2122 \ell^{11}+475 \ell^{12}+56
   \ell^{13}+8 \ell^{14}) a^2 \Omega^2\bigg]+ O\big[(a \Omega)^3\big]
\end{align*}
For our purposes we were required to compute sums for $N=0,...,24$ up to $\Omega^6$, for each of \eqref{eq:samplesumes1}, \eqref{eq:samplesumes2} and \eqref{eq:samplesumes3}.

\section{Expansion of spin-weighted spheroidal harmonic}
\label{sec:spheroidal-expansion}
Using the methods described in Sec.~\ref{sec:spheroidal}, the spin-weighted spheroidal harmonics may be
written as a power series in $a \omega$. In this appendix, we given the explicit form of the
expansion to order $(a \omega)^4$. Defining
\begin{equation}
  \alpha_{\ell} \equiv \frac{1}{\ell}\sqrt{\frac{(\ell ^2-m^2)(\ell ^2-s^2)}{(2 \ell -1)(2 \ell +1)}}
\end{equation}
as in Sec.~\ref{sec:spheroidal} (but without the $s$ and $m$ subscripts for notational compactness),
the expansion of ${}_s S_{\ell m} (\theta, \varphi; a \omega)$ is given by
\begin{align}
{}_s S_{\ell m} & (\theta, \varphi; a \omega) =
  {}_s Y_{\ell m}
  + a \omega  \bigg[
      \frac{s \alpha _{\ell }}{\ell }{}_s Y_{\ell -1, m}
    - \frac{s \alpha _{\ell + 1}}{\ell + 1} {}_s Y_{\ell + 1, m}\bigg]
\nonumber \\ &
  + a^2 \omega ^2 \bigg[
    - \frac{(\ell - 2 s^2) \alpha _{\ell -1} \alpha _{\ell }}{2 \ell(2 \ell-1)} {}_s Y_{\ell -2, m}
    + \frac{m s (\ell ^2 - 2 s^2) \alpha _{\ell }}{\ell ^3 (\ell ^2-1)}{}_s Y_{\ell -1, m} 
    - \bigg(\frac{s^2 \alpha _{\ell }{}^2}{2 \ell ^2}+\frac{s^2 \alpha _{\ell + 1}{}^2}{2 (\ell+1 )^2}\bigg) {}_s Y_{\ell m}
\nonumber \\ & \qquad \qquad
    - \frac{m s (\ell ^2 +2 \ell -2 s^2 + 1) \alpha _{\ell + 1}}{\ell  (\ell+1 )^3 (\ell+2 )} {}_s Y_{\ell + 1, m}
    + \frac{(\ell +2 s^2+1) \alpha _{\ell + 1} \alpha _{\ell + 2}}{2(\ell+1)(2\ell+3)} {}_s Y_{\ell + 2, m} \bigg]
\nonumber \\ &
  + a^3 \omega ^3 \bigg[
      c^{[3,-3]} \, {}_s Y_{\ell -3, m}
    + c^{[3,-2]} \, {}_s Y_{\ell -2, m}
    + c^{[3,-1]} \, {}_s Y_{\ell -1, m}
    + c^{[3,1]} \, {}_s Y_{\ell + 1, m}
    + c^{[3,2]} \, {}_s Y_{\ell + 2, m}
    + c^{[3,3]} \, {}_s Y_{\ell + 3, m}
    \bigg]
\nonumber \\ &
  + a^4 \omega ^4 \bigg[
      c^{[4,-4]} \, {}_s Y_{\ell -4, m}
    + c^{[4,-3]} \, {}_s Y_{\ell -3, m}
    + c^{[4,-2]} \, {}_s Y_{\ell -2, m}
    + c^{[4,-1]} \, {}_s Y_{\ell -1, m}
    + c^{[4,0]} \, {}_s Y_{\ell m}
    + c^{[4,1]} \, {}_s Y_{\ell + 1, m}
\nonumber \\ & \qquad \qquad
    + c^{[4,2]} \, {}_s Y_{\ell + 2, m}
    + c^{[4,3]} \, {}_s Y_{\ell + 3, m}
    + c^{[4,4]} \, {}_s Y_{\ell + 4, m}
    \bigg] + \mathcal{O}\Big[(a\omega)^5\Big],
\end{align}
where the coefficients $c^{[i,j]}$ are given by
\begin{align*}
c^{[3,-3]}&=
  - \frac{s (3 \ell -2 s^2 -1) \alpha _{\ell -2} \alpha _{\ell -1} \alpha _{\ell }}{6 (\ell -1) \ell  (2\ell - 1)} \\
c^{[3,-2]}&=
    \frac{m s^2 (2 \ell ^2+\ell-4 s^2) \alpha _{\ell -1} \alpha _{\ell }}{(\ell -2) \ell ^3 (\ell +1) (2\ell - 1)} \\
c^{[3,-1]}&=
  - \frac{4 m^2 s^3 (\ell^2-s^2 ) \alpha _{\ell }}{(\ell -1)^2 \ell ^5 (\ell +1)^2}
  - \frac{s (3 \ell-2 s^2 -1) \alpha _{\ell -1}^2 \alpha _{\ell }}{2 \ell ^2 (2\ell - 1)}
  - \frac{3 s^3 \alpha _{\ell }^3}{2 \ell ^3}
  + \frac{s (1+2 s^2+3 \ell +s^2 \ell +2 \ell ^2) \alpha _{\ell } \alpha _{\ell +1}^2}{2 \ell ^2 (\ell +1)^2} \\
c^{[3,0]}&=
  - \frac{m s^2 (\ell ^2-2 s^2) \alpha _{\ell }^2}{(\ell -1) \ell ^4 (\ell +1)}
  - \frac{m s^2 (1-2 s^2+2 \ell +\ell ^2) \alpha _{\ell +1}^2}{\ell  (\ell +1)^4 (\ell +2)} \\
c^{[3,1]}&=
    \frac{4 m^2 s^3 (1-s+\ell ) (1+s+\ell ) \alpha _{\ell +1}}{\ell ^2 (\ell +1)^5 (\ell +2)^2}
  + \frac{s (s^2+\ell -s^2 \ell +2 \ell ^2) \alpha _{\ell }^2 \alpha _{\ell +1}}{2 \ell ^2 (\ell +1)^2}
  + \frac{3 s^3 \alpha _{\ell +1}^3}{2 (\ell +1)^3}
  - \frac{s (4+2 s^2+3 \ell ) \alpha _{\ell +1} \alpha _{\ell +2}^2}{2 (\ell +1)^2 (2\ell + 3)} \\
c^{[3,2]}&=
    \frac{m s^2 (1-4 s^2+3 \ell +2 \ell ^2) \alpha _{\ell +1} \alpha _{\ell +2}}{\ell  (\ell +1)^3 (\ell +3) (2\ell + 3)} \\
c^{[3,3]}&=
  - \frac{s (4+2 s^2+3 \ell ) \alpha _{\ell +1} \alpha _{\ell +2} \alpha _{\ell +3}}{6 (\ell +1) (\ell +2) (2\ell + 3)} \\
c^{[4,-4]}&=
    \frac{(8 s^2+4 s^4-3 \ell -12 s^2 \ell +3 \ell ^2) \alpha _{\ell -3} \alpha _{\ell -2} \alpha _{\ell -1} \alpha _{\ell }}{24 (\ell -1) \ell  (2\ell - 3) (2\ell - 1)} \\
c^{[4,-3]}&=
  - \frac{m s (2 s^2+4 s^4-4 s^2 \ell -\ell ^2-2 s^2 \ell ^2+\ell ^3) \alpha _{\ell -2} \alpha _{\ell -1} \alpha _{\ell }}{2 (\ell -3) (\ell -1) \ell ^3 (\ell +1) (2\ell - 1)} \\
c^{[4,-2]}&=
    \frac{m^2 s^2 (16 s^4-4 s^2 \ell -24 s^4 \ell -14 s^2 \ell ^2+12 s^4 \ell ^2+3 \ell ^3+24 s^2 \ell ^3-2 \ell ^4-12 s^2 \ell ^4-\ell ^5+\ell ^6) \alpha _{\ell -1} \alpha _{\ell }}{(\ell -2)^2 (\ell -1)^2 \ell ^5 (\ell +1)^2 (2\ell - 1)}
\\ & \quad
  + \frac{(8 s^2+4 s^4-3 \ell -12 s^2 \ell +3 \ell ^2) \alpha _{\ell -2}^2 \alpha _{\ell -1} \alpha _{\ell }}{12 (\ell -1) \ell  (2\ell - 1)^2}
  + \frac{(2 s^2+4 s^4-8 s^2 \ell +\ell ^2) \alpha _{\ell -1}^3 \alpha _{\ell }}{4 \ell ^2 (2\ell - 1)^2}
\\ & \quad
  - \frac{(-6 s^4+s^2 \ell +16 s^4 \ell -6 s^2 \ell ^2+\ell ^3) \alpha _{\ell -1} \alpha _{\ell }^3}{4 \ell ^3 (2\ell - 1)^2}
\\ & \quad
  + \frac{(10 s^4 \ell -2 (s^2 + 2 s^4) + (8 s^4 + 9 s^2 -1) \ell ^2 + 2 (5 s^2-1) \ell ^3 - \ell ^4) \alpha _{\ell -1} \alpha _{\ell } \alpha _{\ell +1}^2}{4 \ell ^2 (\ell +1)^2 (2\ell - 1)^2} \\
c^{[4,-1]}&=
  - \frac{2 m^3 s^3 (4 s^4-6 s^2 \ell ^2+\ell ^4) \alpha _{\ell }}{(\ell -1)^3 \ell ^7 (\ell +1)^3}
\\ & \quad
  - \frac{m s (-8 s^2-16 s^4+28 s^2 \ell +12 s^4 \ell +\ell ^2-8 s^2 \ell ^2-5 \ell ^3-6 s^2 \ell ^3+3 \ell ^4) \alpha _{\ell -1}^2 \alpha _{\ell }}{2 (\ell -2) (\ell -1) \ell ^4 (\ell +1) (2\ell - 1)}
  - \frac{9 m s^3 (\ell ^2-2 s^2) \alpha _{\ell }^3}{2 (\ell -1) \ell ^5 (\ell +1)}
\\ & \quad
  + \frac{m s (\ell ^2 (\ell +1)^3 (2 \ell + 1) - 2 s^4 (8 + 16 \ell  + 15 \ell ^2 + 3 \ell ^3) + s^2 (-8 - 36 \ell  - 46 \ell ^2 - 18 \ell ^3 + 3 \ell ^4 + 3 \ell ^5)) \alpha_{\ell} \alpha_{\ell+1}^2}{2 (\ell-1) \ell ^4 (\ell+1)^4 (\ell + 2)} \\
c^{[4,0]}&=
  - \frac{m^2 s^2 (12 s^4-12 s^2 \ell ^2+\ell ^4) \alpha _{\ell }^2}{2 (\ell -1)^2 \ell ^6 (\ell +1)^2}
  - \frac{(-4 s^2-8 s^4+20 s^2 \ell +20 s^4 \ell -28 s^2 \ell ^2+\ell ^3) \alpha _{\ell -1}^2 \alpha _{\ell }^2}{8 \ell ^3 (2\ell - 1)^2}
  + \frac{11 s^4 \alpha _{\ell }^4}{8 \ell ^4}
\\ & \quad
  - \frac{m^2 s^2 (1-12 s^2+12 s^4+4 \ell -24 s^2 \ell +6 \ell ^2-12 s^2 \ell ^2+4 \ell ^3+\ell ^4) \alpha _{\ell +1}^2}{2 \ell ^2 (\ell +1)^6 (\ell +2)^2}
\\ & \quad
  - \frac{s^2 (2+4 s^2+8 \ell +5 s^2 \ell +8 \ell ^2+5 s^2 \ell ^2) \alpha _{\ell }^2 \alpha _{\ell +1}^2}{4 \ell ^3 (\ell +1)^3}
  + \frac{11 s^4 \alpha _{\ell +1}^4}{8 (\ell +1)^4}
\\ & \quad
  - \frac{(1+52 s^2+28 s^4+3 \ell +76 s^2 \ell +20 s^4 \ell +3 \ell ^2+28 s^2 \ell ^2+\ell ^3) \alpha _{\ell +1}^2 \alpha _{\ell +2}^2}{8 (\ell +1)^3 (2\ell + 3)^2}
 \\
c^{[4,1]}&=
    \frac{2 m^3 s^3 (1-6 s^2+4 s^4+4 \ell -12 s^2 \ell +6 \ell ^2-6 s^2 \ell ^2+4 \ell ^3+\ell ^4) \alpha _{\ell +1}}{\ell ^3 (\ell +1)^7 (\ell +2)^3}
  + \frac{9 m s^3 (1-2 s^2+2 \ell +\ell ^2) \alpha _{\ell +1}^3}{2 \ell  (\ell +1)^5 (\ell +2)}
\\ & \quad
  + \frac{m s (-8 s^4-5 s^2 \ell -10 s^4 \ell -4 s^2 \ell ^2-12 s^4 \ell ^2+\ell ^3+6 s^4 \ell ^3+4 \ell ^4-12 s^2 \ell ^4+5 \ell ^5-3 s^2 \ell ^5+2 \ell ^6) \alpha _{\ell }^2 \alpha _{\ell +1}}{2 (\ell -1) \ell ^4 (\ell +1)^4 (\ell +2)}
\\ & \quad
\\ & \quad
  - \frac{m s (9-38 s^2-28 s^4+29 \ell -26 s^2 \ell -12 s^4 \ell +34 \ell ^2+10 s^2 \ell ^2+17 \ell ^3+6 s^2 \ell ^3+3 \ell ^4) \alpha _{\ell +1} \alpha _{\ell +2}^2}{2 \ell  (\ell +1)^4 (\ell +2) (\ell +3) (2\ell + 3)} \\
c^{[4,2]}&=
\frac{m^2 s^2 \alpha_{\ell + 1} \alpha_{\ell + 2}}{\ell ^2 (\ell + 1)^5 (\ell +2)^2 (\ell + 3)^2 (2 \ell + 3)} \times \\
\\ &  \qquad
      \Big[52 s^4 -3 - 46 s^2 + 6 (8 s^4 - 24 s^2 -1) \ell  + 2 (6 s^4- 79 s^2 + 2) \ell ^2 + (19 - 72 s^2) \ell ^3 - 6 (2 s^2-3) \ell ^4 + 7 \ell ^5 + \ell ^6 \Big]
\\ & \quad
  - \frac{(3 s^2+6 s^4+12 s^2 \ell -6 s^4 \ell +\ell ^2+21 s^2 \ell ^2-8 s^4 \ell ^2+2 \ell ^3+10 s^2 \ell ^3+\ell ^4) \alpha _{\ell }^2 \alpha _{\ell +1} \alpha _{\ell +2}}{4 \ell ^2 (\ell +1)^2 (2\ell + 3)^2}
\\ & \quad
  - \frac{(1+7 s^2+22 s^4+3 \ell +13 s^2 \ell +16 s^4 \ell +3 \ell ^2+6 s^2 \ell ^2+\ell ^3) \alpha _{\ell +1}^3 \alpha _{\ell +2}}{4 (\ell +1)^3 (2\ell + 3)^2}
\\ & \quad
  + \frac{(1+10 s^2+4 s^4+2 \ell +8 s^2 \ell +\ell ^2) \alpha _{\ell +1} \alpha _{\ell +2}^3}{4 (\ell +1)^2 (2\ell + 3)^2}
  + \frac{(6+20 s^2+4 s^4+9 \ell +12 s^2 \ell +3 \ell ^2) \alpha _{\ell +1} \alpha _{\ell +2} \alpha _{\ell +3}^2}{12 (\ell +1) (\ell +2) (2\ell + 3)^2} \\
c^{[4,3]}&=
  - \frac{m s (2-4 s^2-4 s^4+5 \ell +4 \ell ^2+2 s^2 \ell ^2+\ell ^3) \alpha _{\ell +1} \alpha _{\ell +2} \alpha _{\ell +3}}{2 \ell  (\ell +1)^3 (\ell +2) (\ell +4) (2\ell + 3)} \\
c^{[4,4]}&=
    \frac{(6+20 s^2+4 s^4+9 \ell +12 s^2 \ell +3 \ell ^2) \alpha _{\ell +1} \alpha _{\ell +2} \alpha _{\ell +3} \alpha _{\ell +4}}{24 (\ell +1) (\ell +2) (2\ell + 3) (2\ell + 5)}
\end{align*}
\end{widetext}

\bibliography{MST-PN-Kerr}

\end{document}